\documentclass[%
 reprint,
 amsmath,amssymb,
 aps,prd,
 superscriptaddress,
]{revtex4-2}

\usepackage{booktabs}
\usepackage{graphicx}
\usepackage{bm}
\usepackage{hyperref}
\usepackage{color}

\begin{document}

\def\tov#1{Tolman--Oppenheimer--Volkoff#1 (TOV#1)\gdef\tov{TOV}}
\def\eos#1{equation of state#1 (EOS#1)\gdef\eos{EOS}}
\def\bh#1{black hole#1 (BH#1)\gdef\bh{BH}}
\def\singu#1{\emph{singustar#1}}
\newcommand{\PL}[1]{{\textcolor{red}{\sf{[PL: #1]}}}}

%\graphicspath{{Images/}}

\title{Singustars: Between Black Holes and Compact Stars}

\author{Pablo Laguna}
\affiliation{Center for Gravitational Physics, Weinberg Institute, Department of Physics, The University of Texas at Austin, Austin, TX 78712, USA}

\date{\today}

\begin{abstract}
The Tolman--Oppenheimer--Volkoff equations admit singular solutions in which a negative point mass is surrounded by a self-gravitating fluid. For a $\Gamma$-law polytropic equation of state, these solutions, which we call \emph{singustars}, are thick shells bounded by two zero-pressure surfaces around a naked, repulsive, negative-mass singularity, with a Schwarzschild exterior. They form a two-parameter family, labeled by the mass $M$ and the compactness $C\equiv M/R$. \emph{Singustars} join regular polytropes as the central negative mass vanishes and reach compactness beyond the Buchdahl bound. From exact equilibria and their radial modes, we find that \emph{singustars} with $\Gamma = 2$ are stable only for $M\lesssim0.63$ and $C\lesssim0.25$, a region that nonetheless contains configurations more compact than any stable regular polytrope, although none with a light ring. In the numerically resolved parameter space, unstable \emph{singustars} have a single unstable radial mode; many are energetically unbound. Nonlinear evolutions in ingoing null coordinates show that the end state depends on the perturbation. For the collapsing models studied, backward tracing of outgoing null rays identifies an event-horizon generator with a past endpoint on the singularity, before the first trapped surface appears. Outward perturbations disperse unbound shells and leave the singularity naked, while a bound shell migrates toward, and oscillates about, a stable configuration with the same conserved masses.
\end{abstract}

\maketitle

%=====================================================================
\section{Introduction}
\label{sec:intro}

The \tov{} equations admit singular solutions consisting of a negative point-mass singularity surrounded by a self-gravitating fluid~\cite{anastopoulos}. When the envelope is composed of radiation, such solutions have been proposed as models of a \bh{} firehole in equilibrium with a surrounding bath of Hawking radiation~\cite{Zurek,t_hooft}. Recently, the dynamical stability of \tov{} fireholes with a surrounding fluid obeying a $\Gamma$-law \eos{} with $\Gamma=4/3$ has been investigated~\cite{laguna2026}. For both \bh{} and \tov{} fireholes, no proper surface exists: the envelope engulfing the singularity has infinite extent. This feature makes it difficult to draw firm conclusions about firehole stability. Nonetheless, the results from the study in Ref.~\cite{laguna2026} indicate that \tov{} fireholes are radially unstable unless confined within
approximately their firehole radius.

Here, we introduce a different class of singular solutions in which the surrounding matter is a perfect fluid instead obeying a $\Gamma$-law polytropic \eos{.} As with regular polytropic stars, the configurations possess a well-defined surface and a positive energy density throughout the fluid. At the same time, just as with fireholes, the solutions retain a negative point-mass, naked singularity. They therefore possess some characteristics of \bh{s} and others of compact stars. We refer to such an object as a \singu{}. Because the exterior is Schwarzschild and the fluid occupies a finite shell, the stability problem is well posed, and standard one-dimensional numerical-relativity tools can follow the fate of an unstable \singu{}. \emph{Singustars} also provide a concrete setting in which to ask whether a naked singularity, once surrounded by matter, can be hidden by gravitational collapse.

Our main results are as follows. \emph{Singustars} form a two-parameter family, labeled by mass $M$ and compactness $C\equiv M/R$, that joins the sequence of regular polytropes as the central negative mass vanishes (Sec.~\ref{sec:equilibria}). Radial stability depends on a competition between the repulsive singularity, which stabilizes the inner part of the shell, and the compact outer region, which destabilizes it. \emph{Singustars} are stable in a finite region of low mass and moderate compactness that extends beyond the maximum compactness of stable regular polytropes; the unstable models in the resolved parameter region have one unstable radial mode (Sec.~\ref{sec:stability}). Nonlinear evolutions show that the instability has no unique end state (Sec.~\ref{sec:evolution}). For the collapsing configurations studied, a reconstructed event horizon encloses the negative-mass singularity, outward perturbations of unbound \singu{s} disperse the shell and leave the singularity naked, and the bound unstable model studied migrates toward the stable branch and oscillates about it. We discuss these results in Sec.~\ref{sec:discussion}. We collect technical details in the Appendixes.

Throughout this work, we use units in which $G=c=1$ and denote derivatives with respect to the areal radius $r$ with primes.

%=====================================================================
\section{Singustar equilibria}
\label{sec:equilibria}

\subsection{Field equations and equation of state}

The spacetime metric for a static, spherically symmetric configuration is $ds^2=-e^{2\nu(r)}dt^2+e^{2\lambda(r)}dr^2+r^2\,d\Omega^2$, where $d\Omega^2=d\theta^2+\sin^2\theta\,d\phi^2$. For matter described by a perfect fluid, the Einstein equations reduce to the \tov{} equations~\cite{oppenheimer,Tolman1934-TOLRTA},
\begin{eqnarray}
    m'&=&4\pi r^2\rho\,,\label{eq:tov1a}\\
    p'&=&-(p+\rho)(m+4\,\pi\, p\, r^3)\,e^{2\lambda}/r^2\,,\label{eq:tov2a}
\end{eqnarray}
with $\nu'=-p'/(\rho+p)$ and $e^{-2\lambda}\equiv1-2m/r$. Here $\rho$ denotes the total mass-energy density, $p$ the pressure, and $m(r)$ the gravitational mass enclosed within the areal radius $r$.

We consider the $\Gamma$-law polytropic \eos{} $p = K\,\rho_o^\Gamma$, where $\rho_o$ is the rest-mass density, and $K$ and $n = 1/(\Gamma-1)$ are the polytropic constant and index, respectively. In terms of the total density, $\rho = p/(\Gamma-1)+ (p/K)^{1/\Gamma}$. We take $\Gamma = 2$ and report results for $K=1$; thus, $\rho = p+ \sqrt{p}$. Since for $\Gamma = 2$, $K$ has dimensions of length squared, masses and lengths are in units of $\sqrt{K}$, and dimensionless results depend only on $C$ and $M$. From $\nu'=-p'/(\rho+p)$,
\begin{equation}
e^{2\,\nu}(1+2\,\sqrt{p})^2 = 1-\frac{2M}{R}\,,
\label{eq:firstint}
\end{equation}
where the integration constant is chosen so that at the outer surface $R$, where $p(R) = 0$, the solution matches the exterior Schwarzschild metric of mass $M$. As we shall see, the pressure also vanishes at a radius $R_{\rm in} < R$, so that $e^{2\,\nu(R_{\rm in})}= e^{2\,\nu(R)} = 1-2M/R$. Inside $R_{\rm in}$, the spacetime is negative-mass Schwarzschild with $m(r) = -\mu < 0$, and
\begin{equation}
    e^{2\,\nu}= \left(1+\frac{2\,\mu}{r}\right)\left(1+\frac{2\,\mu}{R_{\rm in}}\right)^{-1}\left( 1-\frac{2\,M}{R}\right)\,.
\end{equation}
A \singu{} is then a thick matter shell surrounding a negative-mass, repulsive, naked singularity.

\subsection{Near-surface structure}

The solution near the two surfaces follows directly from Eqs.~(\ref{eq:tov1a}) and~(\ref{eq:tov2a}). To treat both surfaces on the same footing, we introduce
\begin{equation}
(\mathcal{R},\mathcal{M}) =
\begin{cases}
(R,M), & \text{outer surface},\\[1mm]
(R_{\rm in},-\mu), & \text{inner surface},
\end{cases}
\end{equation}
and write $r=\mathcal{R}(1+\xi)$, with $\xi<0$ inside the outer surface and $\xi>0$ outside the inner surface. To leading order,
\begin{eqnarray}
m &\simeq& \mathcal{M}-\frac{\pi \mathcal{R}^3 \mathcal{M}}{\mathcal{R}-2\mathcal{M}}\,\xi^2\,, \\
p &\simeq& \frac{1}{4}\left(\frac{\mathcal{M}}{\mathcal{R}-2\mathcal{M}}\right)^2\xi^2\,.
\end{eqnarray}
Since $\rho\simeq\sqrt{p}$ as $p\to0$, the density grows linearly with distance from each surface, at rates fixed by the exterior mass $M$ at $R$ and by the central mass $\mu$ at $R_{\rm in}$. These expansions are not needed to construct the equilibria. They determine the behavior of the pulsation coefficients at the two surfaces (Appendix~\ref{app:bc}).

\subsection{The singustar family}

 As $\mu\to0$, the inner radius $R_{\rm in} \rightarrow 0$, and the family joins the sequence of regular $\Gamma=2$ polytropes. For $K=1$ the latter has a maximum mass $M_{\max}=0.1637$ at $C=0.2144$, so for $M<M_{\max}$ there is a band of compactness, bounded by the two regular polytropes of that mass, in which no \singu{} exists (configurations there would require $\mu\le0$).

From Eq.~(\ref{eq:tov2a}), $p'\propto\sqrt{p}$ as $p\to0$. Hence, $p\equiv0$ is also a solution through the surface. This degeneracy is removed by working with $u\equiv\sqrt{p}=\rho_o$, in terms of which Eqs.~(\ref{eq:tov1a})--(\ref{eq:tov2a}) read
\begin{eqnarray}
u' &=& -\frac{(1+2u)(m+4\pi u^2r^3)}{2r(r-2m)}\,,\label{eq:uTOVmain}\\
m' &=& 4\pi r^2(u+u^2)\,.\label{eq:mTOVmain}
\end{eqnarray}
The right-hand sides are smooth in $u$, and at the outer surface $u'(R)=-M/[2R(R-2M)]\neq0$, so $u\equiv0$ is not a solution, and the initial-value problem with $u(R)=0$, $m(R)=M$ has a unique solution. 
We also record the baryon mass
\begin{equation}
M_0 = 4\pi\int_{R_{\rm in}}^{R} \frac{\rho_o\,r^2\,dr}{\sqrt{1-2m/r}}\,,
\label{eq:M0}
\end{equation}
which, together with $\mu$, is conserved by any spherically symmetric evolution of the shell.
Given the exterior parameters $(M,C)$, we integrate Eqs.~(\ref{eq:uTOVmain})--(\ref{eq:mTOVmain}), and the baryon-mass integral Eq.~(\ref{eq:M0}), inward starting at $R = M/C$ with $u(R)=0$, $m(R)=M$, and $M_0=0$ until $u$ vanishes again. 

\begin{figure}[htbp]
    \centering
    \includegraphics[width=\linewidth]{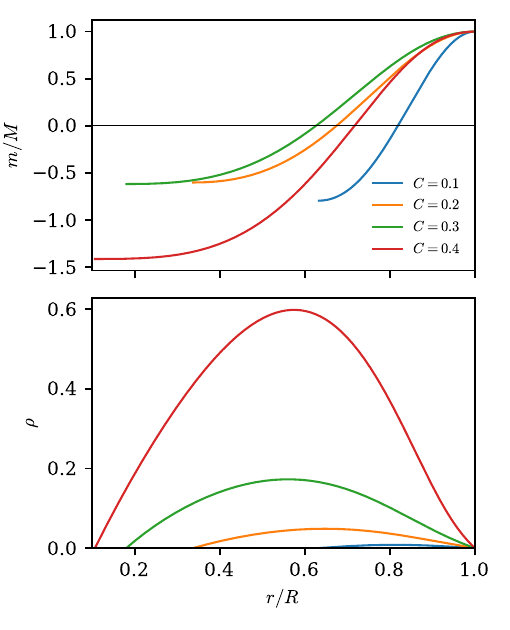}
\caption{Examples of \singu{} solutions for $M=1/3$ and several values of the compactness, as functions of $r/R$. The top panel shows $m/M$, and the bottom panel shows $\rho$.}
\label{fig:Solns_M_1}
\end{figure}

\begin{figure}[htbp]
    \centering
    \includegraphics[width=\linewidth]{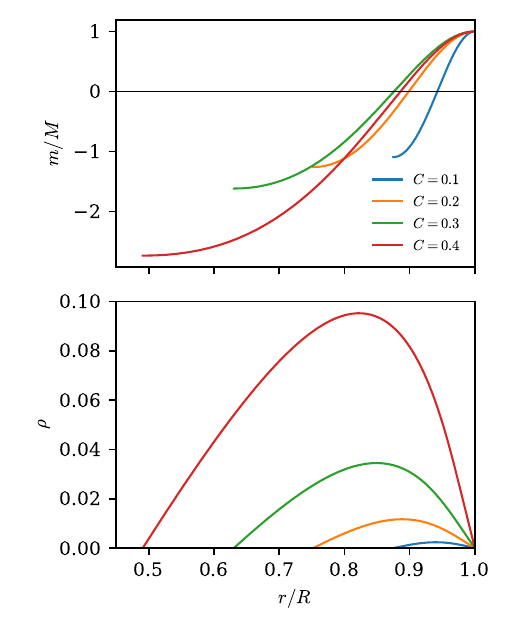}
\caption{Same as Fig.~\ref{fig:Solns_M_1} but for $M=1$.}
\label{fig:Solns_M_10}
\end{figure}

\begin{figure}[htbp]
    \centering
    \includegraphics[width=\linewidth]{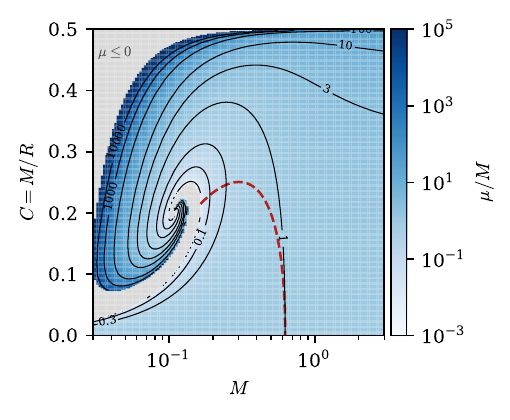}
\caption{Central mass $\mu/M$ as a function of $M$ and compactness $C$. In the gray region, no \singu{} exists ($\mu\le0$); its edge is the sequence of regular $\Gamma=2$ polytropes. The red dashed curve is the marginal-stability curve [solid black curve in Fig.~\ref{fig:stability_map}].}
\label{fig:Solution_Mu_2D}
\end{figure}

\begin{figure}[htbp]
    \centering
    \includegraphics[width=\linewidth]{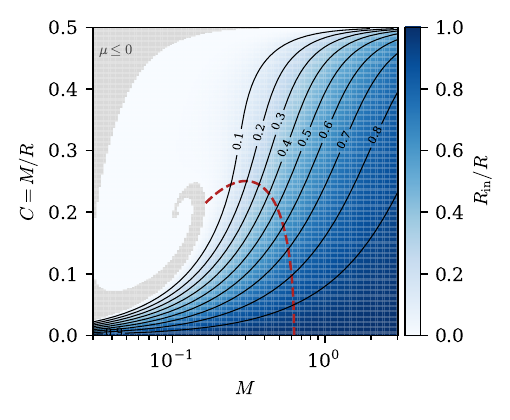}
\caption{Same as Fig.~\ref{fig:Solution_Mu_2D} for the inner radius $R_{\rm in}/R$.}
\label{fig:Solution_Ro_2D}
\end{figure}

Figure~\ref{fig:Solns_M_1} shows examples of \singu{} solutions for $M=1/3$ and several values of the compactness. The top panel shows $m/M$, and the bottom panel shows $\rho$. As compactness increases, the matter shell thickens, and the density peak grows. The central mass, however, is not monotonic in $C$: at $M=1/3$, $\mu/M$ first decreases and then increases with compactness (Table~\ref{tab:modes}). The same trends in thickness and density hold at larger mass, as shown in Fig.~\ref{fig:Solns_M_10} for $M=1$, although the shell is then thinner and the density lower. At $M=1$, by contrast, $\mu$ grows monotonically with $C$. Figures~\ref{fig:Solution_Mu_2D} and~\ref{fig:Solution_Ro_2D} show $\mu/M$ and $R_{\rm in}/R$ across the $(M,C)$ plane.

Two properties of these solutions deserve emphasis. First, \singu{s} exist with $C>4/9$, beyond the Buchdahl bound~\cite{Buchdahl1959}; this does not contradict Buchdahl's theorem, which assumes, among other conditions, a regular stellar interior and an outward nonincreasing density profile. The negative-mass vacuum core and shell profile violate these assumptions. Second, for $C>1/3$ the surface lies inside the photon sphere $r=3M$, so such a \singu{} can trap null geodesics. Light rings, in fact, appear at somewhat lower compactness, inside the matter (Sec.~\ref{sec:stability}). As we show below, however, the configurations with light rings in our resolved parameter space are radially unstable.

%=====================================================================
\section{Radial stability}
\label{sec:stability}

\subsection{Pulsation equation}

To investigate the stability of \singu{s} against radial perturbations, we follow Kokkotas and Ruoff~\cite{kostas_radial}, building on the foundational work of Chandrasekhar~\cite{Chandrasekhar1964a,Chandrasekhar1964b,ChandrasekharTooper1964}. The radial displacement $\delta r(t,r) = \zeta(r)\,e^\nu\,e^{i\,\omega\,t}/r^2$ satisfies
\begin{equation}
\eta' + ({\cal Q} + \omega^2 {\cal W})\zeta = 0\,,\qquad \eta = {\cal P}\,\zeta'\,,
\label{eq:master}
\end{equation}
with
\begin{eqnarray*}
    r^2 {\cal W} &=& (\rho + p)\, e^{3\lambda + \nu}\,,\\
    r^2 {\cal P} &=& \gamma\, p\, e^{\lambda + 3\nu}\,, \\
    r^2 {\cal Q} &=& e^{\lambda + 3\nu} \,(\rho + p)
    \left[ \left(\nu'\right)^2 + \frac{4}{r}\nu' - 8\pi e^{2\lambda} p \right]\,,
\end{eqnarray*}
and $\gamma = (dp/d\rho)(\rho+p)/p=2$ for our $\Gamma = 2$ polytropic \eos{}. Both surfaces are free boundaries at which the Lagrangian pressure perturbation vanishes, $\eta(R_{\rm in})=\eta(R)=0$. For the square-integrable radial modes, these conditions select the regular local solutions at the two limit-point endpoints (Appendix~\ref{app:bc}). By Birkhoff's theorem, the interior $r<R_{\rm in}$ remains negative-mass Schwarzschild with the same $\mu$, so the singularity is a fixed background for radial perturbations and no boundary condition is needed at $r=0$.

\subsection{Variational form and the role of the singularity}

Multiplying Eq.~(\ref{eq:master}) by $\zeta$ and integrating over the shell, where the boundary terms vanish, gives
\begin{equation}
\omega^2 = \frac{\int_{R_{\rm in}}^{R}\left({\cal P}\zeta'^2-{\cal Q}\zeta^2\right)dr}{\int_{R_{\rm in}}^{R}{\cal W}\zeta^2\,dr}\,.
\label{eq:variational}
\end{equation}
For the self-adjoint radial problem defined by the finite-norm endpoint behavior (Appendix~\ref{app:bc}), the eigenvalues are real and ordered, $\omega_0^2<\omega_1^2<\cdots$, and the $n$th eigenfunction has $n$ nodes; the \singu{} is unstable if and only if $\omega_0^2<0$. In a regular star $\nu'>0$, so the term $4\nu'/r$ contributes positively to ${\cal Q}$; this is the general-relativistic destabilization. In a \singu{}, $\nu'<0$ near the inner surface, where the repulsion of the singularity dominates, and ${\cal Q}$ is negative there [Eq.~(\ref{eq:Q1inner})]; the inner part of the shell is therefore stabilizing. Whether a \singu{} is unstable thus depends on the competition between the repulsive inner region and the attractive, compact outer region, and cannot be decided from the sign of ${\cal Q}$ alone. Taking $\zeta={\rm const}$ in Eq.~(\ref{eq:variational}) gives the upper bound $\omega_0^2\le-\int{\cal Q}\,dr/\int{\cal W}\,dr$, so $\int{\cal Q}\,dr>0$ is a sufficient condition for instability; it is satisfied by about one third of the unstable models in Fig.~\ref{fig:stability_map}, all with $M\gtrsim0.63$, and is never violated by the computed spectrum.

\subsection{Numerical method}

We solve the eigenvalue problem in two independent ways. Our primary method discretizes the quadratic forms in Eq.~(\ref{eq:variational}) on a finite-volume grid clustered toward both surfaces, with ${\cal P}=0$ on the two boundary faces, which builds in the free-surface conditions; the result is a symmetric tridiagonal generalized eigenvalue problem. As a check, we also solve the pulsation equations with a shooting method from both surfaces and match the two solutions through their Wronskian at mid-shell. Unlike the equilibrium equations, the pulsation equations are singular at both surfaces, so the shooting starts slightly inside each surface, using the local expansions in Appendix~\ref{app:bc}. The two methods agree to $\sim10^{-6}$ in $\omega^2$, and the finite-volume results converge with resolution. A full resolution study across the near-degenerate, thick-shell limit remains outside the scope of the present calculation. As a test, we take the limit $\mu\to0$ at $M=0.14$, which for $K=100$ corresponds to the $1.4\,M_\odot$, $\Gamma=2$ polytrope commonly used to test numerical-relativity codes. We recover its radius, $R=9.586\,M_\odot$, and its first four radial frequencies, $1.443$, $3.954$, $5.914$, and $7.773$~kHz, in agreement with the perturbative values $1.442$, $3.955$, $5.916$, and $7.776$~kHz~\cite{Font2002}.

\subsection{Stability map}

\begin{figure}[htbp]
    \centering
    \includegraphics[width=\linewidth]{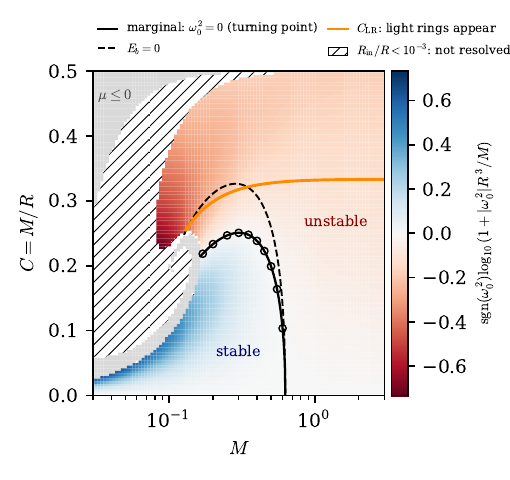}
\caption{Fundamental radial eigenvalue $\omega_0^2$ (in units of $M/R^3$, plotted as ${\rm sgn}(\omega_0^2)\log_{10}(1+|\omega_0^2|R^3/M)$) across the $(M,C)$ plane. Blue: stable; red: unstable. The solid curve is the turning-point condition $\det\,\partial(M_0,\mu)/\partial(M,R)=0$, and the open circles are independent roots of $\omega_0^2=0$. The dashed curve is $E_b\equiv M_0-\mu-M=0$; configurations below it are bound. The orange curve is $C_{\rm LR}(M)$, above which \singu{s} have light rings. Gray: no \singu{} ($\mu\le0$). Hatched: thick shells with $R_{\rm in}/R<10^{-3}$, for which the eigenvalue problem is too ill-conditioned to resolve. }
\label{fig:stability_map}
\end{figure}

Figure~\ref{fig:stability_map} summarizes the stability of the fundamental mode across the $(M,C)$ plane. \emph{Singustars} are stable in a finite region bounded by the regular-polytrope sequence and by a marginal curve $C_{\rm crit}(M)$ that leaves the regular sequence near its maximum-mass configuration, reaches a maximum $C_{\rm crit}\simeq0.251$ at $M\simeq0.30$, and descends to $C\to0$ at $M\simeq0.627$. Stable \singu{s} can therefore be more compact than any stable regular $\Gamma=2$ polytrope ($C\le0.214$). All \singu{s} with $M\gtrsim0.63$, including all the models of Fig.~\ref{fig:Solns_M_10}, are unstable. Of the models in Fig.~\ref{fig:Solns_M_1}, those with $C=0.1$ and $0.2$ are stable and those with $C=0.3$ and $0.4$ are unstable. Nowhere in the numerically resolved region do we find more than one unstable radial mode; no statement about the number of unstable modes is made for the unresolved, highly ill-conditioned configurations.

The stable region also lies entirely below the onset of light rings. Null circular orbits sit where $r\nu'=1$, which by Eq.~(\ref{eq:tov2a}) is $r=3m+4\pi r^3p$. Because $e^{2\nu}/r^2$ diverges at the repulsive singularity and vanishes at infinity, light rings come in pairs, an inner stable one and an outer unstable one~\cite{Cunha2017}. The pair appears at a critical compactness $C_{\rm LR}(M)$ that is below $1/3$, so both light rings can lie inside the matter. For example, $C_{\rm LR}=0.2988$, $0.3175$, $0.3275$, and $0.3319$ for $M=0.2$, $0.3$, $0.5$, and $1$, with $C_{\rm LR}\to1/3$ as $M$ grows. At every mass, $C_{\rm LR}$ exceeds the marginal-stability compactness by at least $\simeq0.06$. Consequently, within the resolved domain, each configuration with a light ring is radially unstable: the stable branch contains no ultracompact \singu{s} with light rings. Table~\ref{tab:modes} lists the equilibrium properties and the three lowest eigenvalues for the models of Figs.~\ref{fig:Solns_M_1} and~\ref{fig:Solns_M_10}. For the unstable models, the unstable mode grows on the dynamical time, $\tau=(-\omega_0^2)^{-1/2}\simeq1.3$--$2.9\,(R^3/M)^{1/2}$, except for the nearly marginal $M=1$, $C=0.1$ model, for which $\tau\simeq4.6\,(R^3/M)^{1/2}$. The overtones are stable in all cases.

\begin{figure}[htbp]
    \centering
    \includegraphics[width=\linewidth]{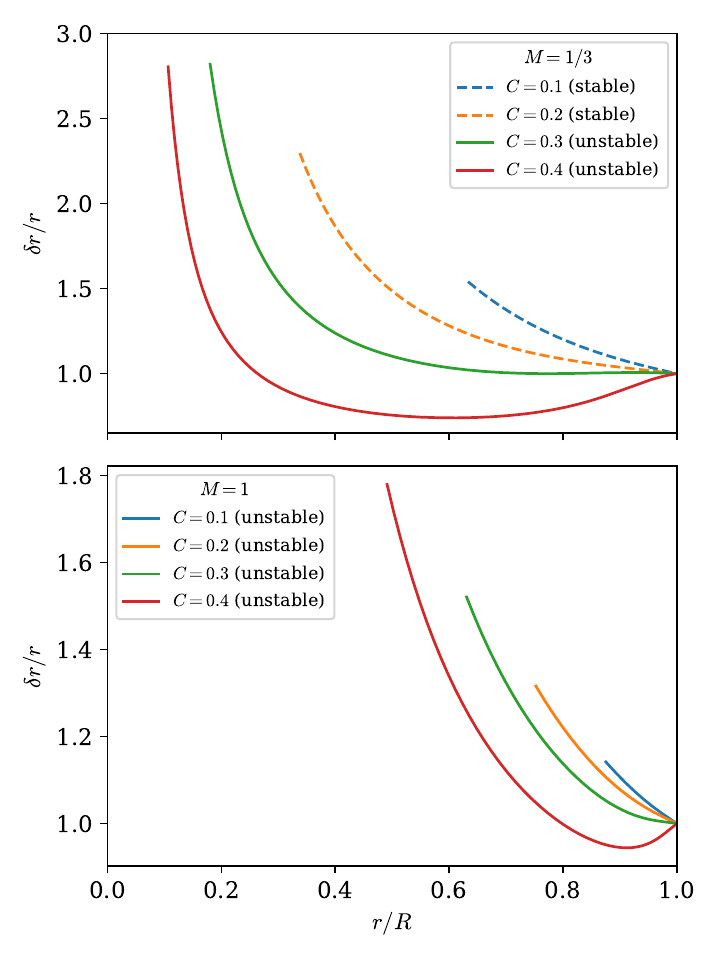}
\caption{Fundamental-mode relative displacement $\delta r/r$, normalized at the outer surface, for the models of Table~\ref{tab:modes}. Top: $M=1/3$; bottom: $M=1$. Dashed: stable; solid: unstable.}
\label{fig:eigenfunctions}
\end{figure}

Figure~\ref{fig:eigenfunctions} shows the fundamental eigenfunctions of the models in Table~\ref{tab:modes}. They are nodeless, and the relative displacement is largest at the inner surface, where it exceeds that at the outer surface by a factor of $1.1$--$2.8$, so the shell motion is not homologous. For the most compact models, it dips below its outer-surface value in the outer part of the shell, by up to $\simeq25\%$.

\begin{table*}[htbp]
\caption{Equilibrium properties and the three lowest radial eigenvalues (in units of $M/R^3$) of the models in Figs.~\ref{fig:Solns_M_1} and~\ref{fig:Solns_M_10}. $E_b\equiv M_0-\mu-M$; $E_b<0$ means the configuration is energetically unbound.}
\label{tab:modes}
\begin{ruledtabular}
\begin{tabular}{ccccccccc}
$M$ & $C$ & $R_{\rm in}/R$ & $\mu/M$ & $M_0/M$ & $E_b/M$ & $\omega_0^2$ & $\omega_1^2$ & $\omega_2^2$ \\
\colrule
1/3 & 0.1 & 0.6331 & 0.795 & 1.806 & $+0.011$ & $+0.1328$ & 15.61 & 35.76 \\
1/3 & 0.2 & 0.3376 & 0.602 & 1.645 & $+0.043$ & $+0.1657$ & 8.984 & 20.12 \\
1/3 & 0.3 & 0.1802 & 0.619 & 1.650 & $+0.032$ & $-0.2580$ & 5.826 & 13.34 \\
1/3 & 0.4 & 0.1064 & 1.415 & 1.963 & $-0.452$ & $-0.6080$ & 2.959 & 7.117 \\
1 & 0.1 & 0.8750 & 1.094 & 2.089 & $-0.005$ & $-0.0466$ & 40.13 & 92.43 \\
1 & 0.2 & 0.7521 & 1.260 & 2.226 & $-0.034$ & $-0.1230$ & 18.75 & 43.01 \\
1 & 0.3 & 0.6305 & 1.617 & 2.474 & $-0.143$ & $-0.2453$ & 11.19 & 25.41 \\
1 & 0.4 & 0.4909 & 2.735 & 3.046 & $-0.689$ & $-0.4064$ & 6.738 & 15.06 \\
\end{tabular}
\end{ruledtabular}
\end{table*}

\subsection{Turning-point criterion}

The marginal curve can be identified independently of the eigenvalue calculation. A spherical radial perturbation conserves the core mass parameter $\mu$ and the baryon mass $M_0$. Let $q^A=(M,R)$ parameterize the equilibria and $I^a(q)=(M_0(q),\mu(q))$ their conserved quantities. A nonzero tangent $\delta q^A$ preserving both invariants to first order must satisfy
\begin{equation}
\frac{\partial I^a}{\partial q^A}\,\delta q^A=0\,.
\end{equation}
Such a direction can exist only when
\begin{equation}
\det\frac{\partial(M_0,\mu)}{\partial(M,R)}=0\,.
\label{eq:turningjac}
\end{equation}
Differentiating the equilibrium equations along this direction gives a time-independent solution of the linearized interior equations. When the induced displacement preserves baryon number and the core mass and satisfies the two free-surface conditions, it represents an admissible $\omega^2=0$ radial mode. This is the natural two-parameter analog of the stellar turning-point argument~\cite{Sorkin1982,FriedmanIpserSorkin1988}; the Jacobian condition alone need not establish which spectral branch crosses zero. Along fixed-$M$ sequences, the independently calculated fundamental-mode roots and the roots of Eq.~(\ref{eq:turningjac}) agree to $\sim10^{-7}$ in $C$ (Fig.~\ref{fig:stability_map}), providing a strong numerical check of this identification.

\subsection{Binding energy and candidate end states}

The binding energy $E_b\equiv M_0-\mu-M$ compares the \singu{} with its constituents dispersed to infinity: a cold shell at rest far from the singularity with total mass $M_0-\mu$. If $E_b>0$, complete dispersal into a cold shell at rest at infinity is energetically excluded under conservation of $M_0$ and $\mu$. This energetic criterion is distinct from radial mode stability: $E_b>0$ does not imply $\omega_0^2>0$, and $E_b<0$ permits but does not force dispersal. The bound region (below the dashed curve in Fig.~\ref{fig:stability_map}) contains the entire stable region and extends beyond it, up to $C\simeq0.326$. Unstable but bound \singu{s} exist, such as $M=1/3$, $C=0.3$ (Table~\ref{tab:modes}). For $M\gtrsim0.63$, by contrast, \singu{s} are both unstable and unbound, so the shell can fully disperse. This identifies the candidate end states of the instability: expansion and dispersal of the shell, collapse of the shell within $r=2M$, with the formation of an event horizon around the singularity, or, close to the marginal curve, migration to a stable \singu{} with the same $M_0$ and $\mu$. We test these possibilities with nonlinear evolutions in Sec.~\ref{sec:evolution}.

Our analysis does not resolve the hatched region of Fig.~\ref{fig:stability_map}. These are thick shells at small mass, adjacent to the unstable branch of regular polytropes, in which $R_{\rm in}/R<10^{-3}$, the singularity is extremely strong ($\mu/R_{\rm in}\gtrsim10^{6}$), and ${\cal Q}$ varies over $\sim30$ orders of magnitude across the shell; the eigenvalue problem is then not reliably solvable in double precision.

%=====================================================================
\section{Nonlinear evolution}
\label{sec:evolution}

\subsection{Evolution code}
\label{sec:ef}

We evolve the \singu{s} on ingoing null slices, using ingoing Eddington--Finkelstein (Bondi-type) coordinates,
\begin{equation}
ds^2=-e^{2\beta}F\,dv^2+2e^{\beta}dv\,dr+r^2d\Omega^2\,,
\label{eq:efmetric}
\end{equation}
where $F=1-2\,m/r$, $m$ is the Misner--Sharp mass~\cite{MisnerSharp1964}, and $v$ is the advanced time. The hypersurfaces $v={\rm const}$ are ingoing null hypersurfaces, rather than spacelike slices; these coordinates are regular at future horizons, allowing one code to follow dispersal, oscillation, and collapse through apparent-horizon formation. We set $\beta=0$ at the outer surface, so that $v$ is the advanced time of the exterior Schwarzschild metric; inside the cavity $r<R_{\rm in}$, $m=-\mu$ and the metric is negative-mass Schwarzschild in the same form. With $z\equiv u_r=e^\beta u^v$, each fluid element moves according to $dR/dv=e^\beta U/z$, where $U=(Fz^2-1)/(2z)$ is the Misner--Sharp velocity. With $Y\equiv(e+p)z^2$ and $e=\rho_o(1+\epsilon)$ the total energy density ($e\equiv\rho$ of Secs.~\ref{sec:equilibria} and~\ref{sec:stability}), the hypersurface equation for $\beta$ is $\partial_r\beta=4\pi rY$, and the momentum of a zone moving with the fluid obeys
\begin{equation}
\frac{d}{dv}\int r^2Y\,dr=-\left[e^\beta r^2p\right]+\int e^\beta\left(2pr-Ym\right)dr\,,
\label{eq:efmomentum}
\end{equation}
which reduces to the \tov{} equation for a static configuration. For $\Gamma=2$ the velocity follows algebraically from the evolved variables, so no implicit solve is needed. Appendix~\ref{app:ef} gives the derivation and the discretization. The grid is Lagrangian, labeled by the enclosed baryon mass, and both surfaces are free, so no artificial atmosphere is needed and the baryon mass $M_0$ is conserved exactly. The region inside the shell remains negative-mass Schwarzschild with the same $\mu$ (Birkhoff's theorem), so the singularity is never on the grid and acts on the fluid only through the value $m=-\mu$ at the inner surface. Initial data are the exact equilibria of Sec.~\ref{sec:equilibria} with a velocity perturbation of amplitude $|U|_{\max}=10^{-4}$ along the fundamental eigenfunction, inward or outward.

\subsection{Convergence tests}
\label{sec:linear}

We test the code in the linear regime, where the exact answer is known from Sec.~\ref{sec:stability}, and through the Hamiltonian constraint $\partial_rm=4\pi r^2X$ (Appendix~\ref{app:ef}), which is not enforced during the evolution. Table~\ref{tab:convergence} summarizes the results for the unstable model $M=1$, $C=0.30$. The outer-surface radius grows exponentially at a rate that converges to $\sigma_0=\sqrt{-\omega_0^2}$. Measured while $10^{-4}<|R_{\rm out}-R|/R<10^{-2}$, after the initial transient has decayed and before nonlinear effects set in, the relative error falls from $6.8\%$ at $N=200$ to $0.31\%$ at $N=1600$, at an observed order of $1.6$--$1.7$ up to $N=800$ and $1.1$ between $N=800$ and $1600$. At $N=800$ the growth rates of $M=1$, $C=0.45$ and $M=10$, $C=0.30$ agree with $\sigma_0$ to $0.7\%$ and $0.6\%$. The stable model $M=0.3$, $C=0.15$ oscillates at $\omega_0$ to within $1.7\times10^{-4}$, $8.9\times10^{-5}$, and $5.6\times10^{-5}$ for $N=200$, $400$, and $800$. The Hamiltonian-constraint residual converges at second order, decreasing by a factor of $4.0$ with each doubling of $N$, in both the stable and the unstable runs. In the nonlinear phase, it remains below $1.3\times10^{-4}$ in the dispersal and oscillation runs, and below $10^{-3}$ in the collapse runs until shortly after apparent-horizon formation (Table~\ref{tab:horizons}). The time to apparent-horizon formation, the radius at which the trapped region first appears, and the inner-surface radius at that time agree to better than $0.5\%$ between $N=400$ and $800$.

\begin{table}[htbp]
\caption{Convergence tests for $M=1$, $C=0.30$. $\delta\sigma/\sigma_0$ is the relative error of the linear growth rate, measured while $10^{-4}<|R_{\rm out}-R|/R<10^{-2}$; $\|{\cal H}\|$ is the maximum, while $|R_{\rm out}-R|/R<10^{-3}$, of $\sum_j|\Delta m_j-4\pi\int_j r^2X\,dr|/M$, the discrete $L_1$ norm of the Hamiltonian-constraint residual; $\Delta v_{\rm AH}$ is the advanced time from $|R_{\rm out}-R|/R=10^{-2}$ to apparent-horizon formation; $R_{\rm in}/\mu$ and $r_{\rm AH}/M$ are evaluated at apparent-horizon formation. The linear tests use an inward kick $|U|_{\max}=10^{-6}$, the collapse runs $|U|_{\max}=10^{-4}$.}
\label{tab:convergence}
\setlength{\tabcolsep}{3pt}
\begin{ruledtabular}
\begin{tabular}{cccccc}
$N$ & $\delta\sigma/\sigma_0$ & $\|{\cal H}\|$ & $\dfrac{\Delta v_{\rm AH}}{M}$ & $\dfrac{R_{\rm in}}{\mu}$ & $\dfrac{r_{\rm AH}}{M}$ \\[6pt]
\colrule
200 & $0.068$ & $1.4\times10^{-7}$ & 39.82 & 0.305 & 1.981 \\
400 & $0.021$ & $3.5\times10^{-8}$ & 39.86 & 0.306 & 1.974 \\
800 & $0.0067$ & $8.8\times10^{-9}$ & 39.86 & 0.306 & 1.975 \\
1600 & $0.0031$ & $2.2\times10^{-9}$ & --- & --- & --- \\
\end{tabular}
\end{ruledtabular}
\end{table}

\subsection{End states}

\begin{figure*}[htbp]
    \centering
    \includegraphics[width=\textwidth]{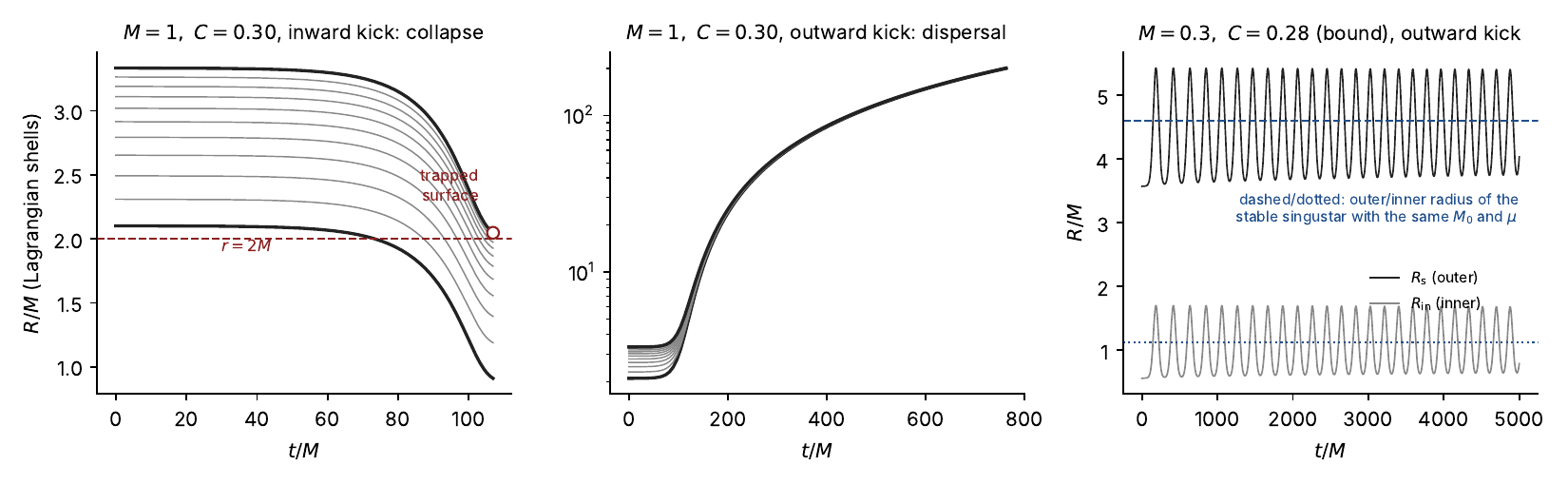}
\caption{Worldlines of Lagrangian shells for the unstable model $M=1$, $C=0.30$ ($|U|_{\max}=10^{-4}$). Left: perturbed inward ($N=800$), the shell collapses; a trapped surface first appears in the fluid at the circle, and the evolution is shown while the Hamiltonian constraint holds to $10^{-3}$. Right: perturbed outward ($N=400$), the unbound shell disperses; the run ends when the inner surface reaches $60\,R(0)$.}
\label{fig:endstates}
\end{figure*}

Unstable \singu{s} collapse when kicked inward and disperse when kicked outward, unless they are bound. Figure~\ref{fig:endstates} shows both outcomes for $M=1$, $C=0.30$. With an inward perturbation, the inner layers are first pushed toward the singularity, and a trapped surface ($2m/r\to1$) forms when the outer layers reach $r\simeq2M$, $39.86M$ in advanced time after the outer surface has departed from equilibrium by $1\%$. We follow the collapse beyond this point in Sec.~\ref{sec:horizon}. With an outward perturbation, the unbound shell ($E_b<0$) expands with nearly uniform velocity, essentially without shock heating, and its asymptotic kinetic energy approaches $-E_b$: the kinetic energy $\sum\Delta A\,(\sqrt{1+U^2}-1)$ when the inner surface reaches $60\,R(0)$ is $0.140\,M$, compared with $-E_b=0.143\,M$. Dispersal leaves behind an isolated negative-mass, naked singularity. The same two outcomes occur for every unbound model we evolved: for $M=1$, $C=0.45$ and $M=10$, $C=0.30$ the dispersing shells carry kinetic energies of $2.07\,M$ and $0.304\,M$, compared with $-E_b=2.11\,M$ and $0.314\,M$, the latter while the shell is still at $59\,R(0)$.

\begin{figure}[htbp]
    \centering
    \includegraphics[width=\linewidth]{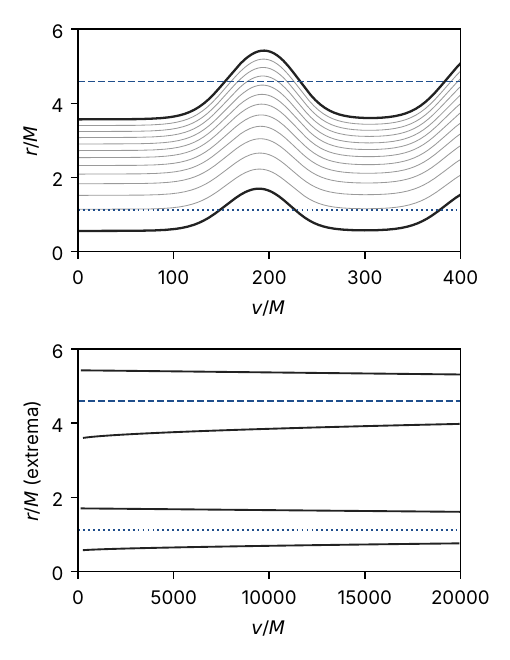}
\caption{The unstable but bound model $M=0.3$, $C=0.28$ perturbed outward ($|U|_{\max}=10^{-4}$, $N=400$) oscillates about the stable \singu{} with the same $M_0$ and $\mu$ ($C=0.217$), whose outer and inner radii are shown by the dashed and dotted lines. Top: worldlines of Lagrangian shells over the first two oscillations. Bottom: successive maxima and minima of the outer (upper pair) and inner (lower pair) surface radii over $2\times10^4M$.}
\label{fig:oscillation}
\end{figure}

Bound \singu{s} ($E_b>0$) cannot disperse. For the unstable but bound model $M=0.3$, $C=0.28$, an inward perturbation again leads to collapse, but an outward perturbation produces large nonlinear oscillations around the location of the stable \singu{} with the same $M_0$ and $\mu$ ($C=0.217$, $R=4.60M$; Fig.~\ref{fig:oscillation}), the analog of the migration of unstable neutron stars to the stable branch~\cite{Font2002}. The period is $\simeq172M$ in advanced time. The outer surface initially swings between $3.57M$ and $5.43M$, and after $2\times10^4M$ between $3.98M$ and $5.32M$. Damping is slow because the flow develops no strong shocks. The agreement of the oscillation center with the conserved-$(M_0,\mu)$ stable equilibrium supports migration, but persistent oscillations do not by themselves demonstrate asymptotic settling.

\subsection{Horizon formation}
\label{sec:horizon}

\begin{figure}[htbp]
    \centering
    \includegraphics[width=\linewidth]{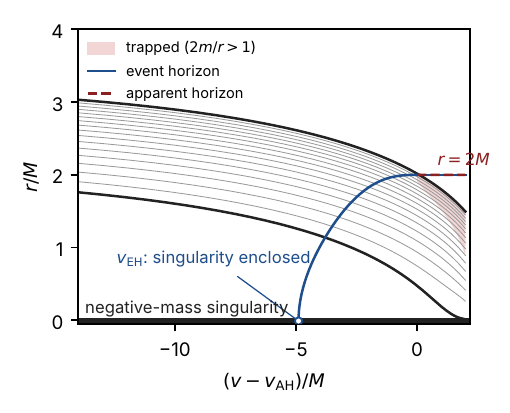}
\caption{Collapse of $M=1$, $C=0.30$ in ingoing Eddington--Finkelstein coordinates ($N=800$). Gray: worldlines of Lagrangian shells; thick black: the two surfaces and the negative-mass singularity at $r=0$. Shaded: trapped region ($2m/r>1$). The reconstructed event-horizon generator (solid) has a past endpoint at the singularity at $v_{\rm EH}$ and joins $r=2M$ when the outer surface crosses it; thereafter, the apparent horizon (dashed) coincides with $r=2M$.}
\label{fig:ef_spacetime}
\end{figure}

Figure~\ref{fig:ef_spacetime} shows the collapse of $M=1$, $C=0.30$. A trapped region first appears inside the fluid, at $r\simeq1.97M$, and the apparent horizon reaches $r=2M$ as the outer surface crosses it, $0.16M$ later in $v$. We trace an outgoing null ray, $dr/dv=\frac{1}{2}e^\beta F$, backward from the late-time exterior horizon at $r=2M$, after the surface has crossed it. In the vacuum exterior, this null generator remains at $r=2M$; its backward continuation across the moving shell is used to reconstruct the boundary of rays able to escape to future null infinity. This event-horizon identification assumes the exterior settles to Schwarzschild and that the subsequent, unresolved interior evolution does not alter the exterior causal structure. Inside the cavity $r<R_{\rm in}$, $F=1+2\mu/r>0$, so outgoing rays emanate from the singularity, and the reconstructed event-horizon generator has its past endpoint there at advanced time $v_{\rm EH}$, $4.82M$ before the apparent horizon forms (at both $N=400$ and $800$). Under the stated late-time continuation, outgoing signals from the singularity after $v_{\rm EH}$ cannot reach future null infinity. The agreement of the reported $v_{\rm AH}-v_{\rm EH}$ at $N=400$ and 800 provides a resolution check, but a separate sensitivity study of the backward integration endpoint would further strengthen the global interpretation. Table~\ref{tab:horizons} gives the results for the other models.

\begin{table}[htbp]
\caption{Horizon formation in the horizon-penetrating evolutions ($N=400$ unless noted; $|U|_{\max}=10^{-4}$, inward). $r_{\rm AH}$ is the radius at which the trapped region first appears, $v_{\rm AH}-v_{\rm EH}$ the advanced time between the enclosure of the singularity by the event horizon and apparent-horizon formation, $R_{\rm in}/\mu$ the inner-surface radius at $v_{\rm AH}$, and $\Delta v_{\rm end}$ the advanced time after $v_{\rm AH}$ during which the Hamiltonian constraint holds to $10^{-3}$. $^{a}N=800$.}
\label{tab:horizons}
\setlength{\tabcolsep}{3pt}
\begin{ruledtabular}
\begin{tabular}{cccccc}
$M$ & $C$ & $\dfrac{r_{\rm AH}}{M}$ & $\dfrac{v_{\rm AH}-v_{\rm EH}}{M}$ & $\dfrac{R_{\rm in}}{\mu}$ & $\dfrac{\Delta v_{\rm end}}{M}$\\[6pt]
\colrule
1 & 0.30 & 1.974 & 4.82 & 0.306 & 1.97 \\
1$^{a}$ & 0.30 & 1.975 & 4.82 & 0.306 & 2.03 \\
1 & 0.45 & 1.993 & 7.40 & 0.100 & 3.84 \\
10 & 0.30 & 2.000 & 2.37 & 0.760 & 3.30 \\
\end{tabular}
\end{ruledtabular}
\end{table}

Inside the horizon, the outer layers are trapped, but the innermost layers remain untrapped because $m=-\mu$ at the inner surface. The compression does work on the inner layers ($dm/dv>0$ for infall), so the trapped region spreads inward; within $2M$ in advanced time after $v_{\rm AH}$, it contains two thirds of the baryon mass. Meanwhile, the shell drives the inner layers onto the repulsive core: $R_{\rm in}$ falls from $0.31\mu$ to $\simeq0.01\mu$, and the peak density rises by a factor of $\simeq40$. This phase is converged for $\simeq1.5M$, and to $\simeq20\%$ in $R_{\rm in}$ up to $2M$; beyond it, the inner layers are unresolved, and the evolution ends. Whether they bounce off the core or the fluid ends in a spacelike singularity remains open. For the bound model $M=0.3$, $C=0.28$, a trapped surface forms at $r\simeq1.5M$ while the outer surface is still at $2.4M$, but the inner layers ($r\simeq0.02\mu$) are already unresolved, so we cannot trace its event horizon.

%=====================================================================
\section{Discussion}
\label{sec:discussion}

We have introduced \singu{s}, singular solutions of the \tov{} equations in which a $\Gamma=2$ polytropic shell with two free surfaces surrounds a repulsive, negative-mass naked singularity. Unlike fireholes, \singu{s} have a finite extent and a Schwarzschild exterior, which makes their stability analysis well posed. They form a two-parameter family that connects continuously to regular polytropes as $\mu\to0$, and they exist with compactness beyond the Buchdahl bound and inside the photon sphere. Exact equilibria are obtained by a single inward integration of the \tov{} equations written in terms of $\sqrt{p}$.

The radial stability of \singu{s} is not decided by the singularity alone. The repulsive inner region stabilizes the shell, and the compact outer region destabilizes it. \emph{Singustars} are stable in a finite region of the $(M,C)$ plane, $M\lesssim0.63$ and $C\lesssim0.25$, bounded by a marginal curve that coincides with the turning point of the map $(M,R)\mapsto(M_0,\mu)$. Within it, stable \singu{s} can exceed the compactness of the most compact stable regular polytrope, but none has a light ring. In the resolved domain the higher-mass configurations are unstable, with one unstable mode growing on the dynamical time; many are also energetically unbound.

\begin{figure*}[htbp]
    \centering
    \includegraphics[width=\textwidth]{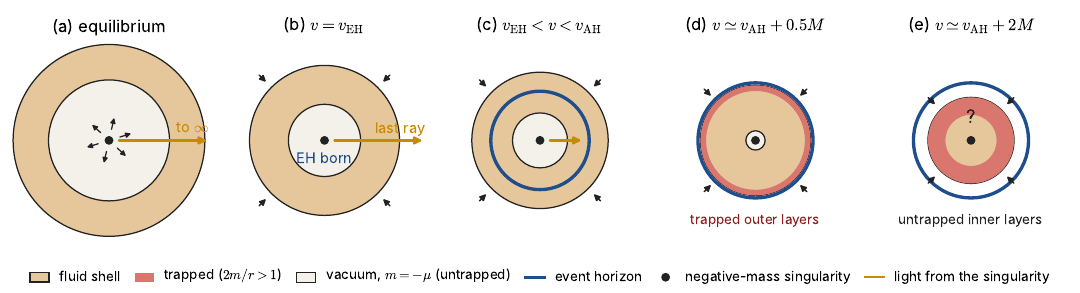}
\caption{Schematic of the collapse of an unstable \singu{} perturbed inward, drawn as equatorial cross sections with radii approximately to scale from the run $M=1$, $C=0.30$ of Fig.~\ref{fig:ef_spacetime}. (a) Equilibrium: the shell surrounds a naked, repulsive negative-mass singularity, whose light reaches infinity. (b) At $v=v_{\rm EH}$, the reconstructed event-horizon generator has a past endpoint at the singularity; the ray shown is the last to escape. (c) The event horizon grows through the collapsing shell; light from the singularity can no longer escape, although no trapped surface has formed. (d) After the outer surface crosses $r=2M$, the event horizon sits at $r=2M$, and the outer layers of the shell are trapped. (e) The trapped region spreads inward while the inner layers, driven onto the repulsive core, remain untrapped; their fate is not determined by our evolutions.}
\label{fig:cartoon}
\end{figure*}

The nonlinear evolutions show that the instability does not select a single end state. In the collapsing runs, the reconstructed event horizon encloses the negative-mass singularity. Figure~\ref{fig:cartoon} sketches the sequence. As the shell collapses, the reconstructed horizon generator has a past endpoint at the singularity at $v_{\rm EH}$, several $M$ in advanced time before the apparent horizon forms, so the singularity is hidden from infinity well before any trapped surface appears [Fig.~\ref{fig:cartoon}(b,c)]. The horizon then grows through the shell and reaches $r=2M$ as the outer surface crosses it. In this collapsing branch, an event horizon encloses the initially naked singularity. Because the initial data already contain a naked singularity, this result bears on, but does not test, standard formulations of cosmic censorship. The resulting black-hole region, the part of spacetime that cannot communicate with infinity, differs from the interior of a Schwarzschild black hole, however. The negative-mass singularity is timelike, and the region around it remains untrapped, since $2m/r=-2\mu/r<0$ between the singularity and the inner surface of the shell. Trapped surfaces form only in the infalling fluid, starting in its outer layers and spreading inward [Fig.~\ref{fig:cartoon}(d,e)]. Whether the interior eventually develops a spacelike singularity, as in ordinary gravitational collapse, is not determined by our evolutions.

Outward perturbations of unbound \singu{s} disperse the shell, leaving behind an isolated negative-mass naked singularity, so the instability does not in general remove the naked singularity. The bound unstable model tested migrates toward the stable branch and undergoes persistent oscillations around it. Which outcome occurs depends on the sign of the perturbation, not only on the equilibrium.

Several limitations remain. All results are for $\Gamma=2$ and a single perturbation profile; we expect the qualitative picture to hold for other stiff equations of state, but this should be checked. The evolution inside the event horizon ends when the inner layers, driven onto the repulsive core, can no longer be resolved; regridding near the inner surface or excising the innermost trapped layers would allow it to continue. The thick-shell region of Fig.~\ref{fig:stability_map} with $R_{\rm in}/R<10^{-3}$ is not resolved by our eigenvalue solver. In addition, nonradial perturbations, unlike radial ones, reach the singularity. In the tortoise coordinate, the singularity lies at a finite distance, and the master potential for a field of spin $s$ approaches $(s^2-1)/(4x^2)$ there, where $x$ is the tortoise distance from the singularity. Axial gravitational perturbations ($s=2$) are then in the limit-point case and evolve uniquely, with no boundary condition. Scalar and electromagnetic perturbations are in the limit-circle case and require one, a choice of self-adjoint extension of the spatial operator~\cite{Wald1980,IshibashiWald2003,GibbonsHartnollIshibashi2005}. We analyze geodesics and nonradial perturbations of \singu{s} in a forthcoming paper.

A further limitation, and the main obstacle to studying \singu{s}
beyond spherical symmetry, is that we have not been able to evolve them with
standard three-dimensional numerical-relativity methods. Our one-dimensional
codes avoid the singularity altogether: the grid is Lagrangian, the region
$r<R_{\rm in}$ is negative-mass Schwarzschild by Birkhoff's theorem, and the
singularity enters only through the boundary value $m=-\mu$ at the inner
surface. None of this carries over to a 3+1 evolution on a Cartesian grid.
The moving-puncture approach, which handles black-hole singularities without
excision, relies on the singularity being hidden behind a horizon. In
puncture data the conformal factor is $\psi = 1 + M/(2\bar r)$, so the
BSSN variable $\chi=\psi^{-4}$ vanishes smoothly at the puncture, and the
$1+\log$ slicing and $\Gamma$-driver shift then settle into a trumpet
geometry in which errors generated near the puncture stay inside the
horizon. For a \singu{}, none of these ingredients is available.
With $M\to-\mu$ the conformal factor $\psi = 1-\mu/(2\bar r)$ vanishes at
the finite coordinate sphere $\bar r=\mu/2$, where the areal radius goes to
zero, so $\chi$ diverges on a two-sphere instead of vanishing at a point.
The static lapse, $e^{\nu}\propto(1+2\mu/r)^{1/2}$, grows without bound
toward the singularity instead of collapsing. Most importantly, there is no
horizon: the singularity is timelike and in causal contact with the shell
and with infinity, so numerical errors generated there are not confined.
Excision does not resolve this either, because the region around the
singularity is untrapped ($2m/r<0$). An excision surface there would be
timelike, with characteristics entering the computational domain, and would
need boundary data. In spherical symmetry Birkhoff's theorem supplies this
data. Beyond spherical symmetry it does not, which is precisely the boundary
condition that nonradial perturbations, and ultimately binary
\singu{s}, would require. Finding a gauge and boundary treatment
that keeps a naked, repulsive singularity stable in a 3+1 evolution, or a
non-singular source that reproduces its effect on the shell, is the key
step toward simulating \singu{} mergers.

\begin{acknowledgments}
This work is supported by NSF grants PHY-2411068 and PHY-2207780. The author used Claude (Anthropic; Claude Opus 5.5) to help draft and revise parts of the manuscript and check derivations. The same tools were used to produce the data and the figures. The author steered the choice of formulations, methods, and implementations, as well as the validation of the tools. The author takes full responsibility for all the content in the study. 
\end{acknowledgments}

%=====================================================================
\appendix

\section{Boundary conditions for radial perturbations}
\label{app:bc}

The radial perturbations satisfy
\begin{align}
    \eta' + \left({\cal Q}+\omega^2{\cal W}\right)\zeta &=0,
    \label{eq:pert_eta}\\
    \eta &= {\cal P}\zeta',
    \label{eq:pert_zeta}
\end{align}
with ${\cal W}$, ${\cal P}$, and ${\cal Q}$ as in Sec.~\ref{sec:stability}. In contrast to the equilibrium equations, both endpoints of the fluid shell are regular singular points of the pulsation equations, because ${\cal P}\propto(\text{distance})^2\to0$ there, while ${\cal Q},{\cal W}\propto\text{distance}$. Each surface admits one regular and one singular local solution. The finite-volume method selects the regular solution automatically; the shooting method needs the local expansions derived below to start on the regular branch slightly away from each surface. We establish the spectral endpoint classification after the expansions.

\subsection{Outer boundary}

Let $s\equiv R-r>0$. From the equilibrium surface expansion,
\begin{eqnarray}
    p &=& \frac{M^2}{4R^2(R-2M)^2}s^2+O(s^3),\nonumber\\
    \rho+p &=& \frac{M}{2R(R-2M)}s+O(s^2),
\end{eqnarray}
so that ${\cal P} = P_2^{(+)}s^2$, ${\cal W} = W_1^{(+)}s$, and ${\cal Q} = Q_1^{(+)}s$ to leading order, with
\begin{eqnarray}
    P_2^{(+)} &=& \frac{M^2}{2R^5(R-2M)},\nonumber\\
    W_1^{(+)} &=& \frac{M}{2R^2(R-2M)^2},\nonumber\\
    Q_1^{(+)} &=& \frac{M^2(4R-7M)}{2R^6(R-2M)^2}.
    \label{eq:outercoeffs}
\end{eqnarray}
The regular solution is $\zeta = \zeta_R+\zeta_1^{(+)}s+O(s^2)$, $\eta = \eta_2^{(+)}s^2+O(s^3)$, with
\begin{eqnarray}
    \eta_2^{(+)} &=& \frac{1}{2}\left(Q_1^{(+)}+\omega^2W_1^{(+)}\right)\zeta_R ,\nonumber\\
    \zeta_1^{(+)} &=& -\frac{Q_1^{(+)}+\omega^2W_1^{(+)}}{2P_2^{(+)}}\zeta_R .
\end{eqnarray}
It automatically satisfies the free-surface condition $\eta(R)=0$. With $\zeta_R=1$, the integration is started at $r=R-\delta$ using
\begin{align}
    \zeta(R-\delta) &= 1-\frac{\delta}{2(R-2M)}\left[\frac{4R-7M}{R}+\frac{\omega^2R^3}{M}\right]\nonumber\\
    &\quad+O(\delta^2),
    \label{eq:zeta_outer_bc}\\
    \eta(R-\delta) &= \frac{1}{2}\left(Q_1^{(+)}+\omega^2W_1^{(+)}\right)\delta^2+O(\delta^3).
    \label{eq:eta_outer_bc}
\end{align}

\subsection{Inner boundary}

Let $x\equiv r-R_{\rm in}>0$. The equilibrium expansion at $R_{\rm in}$ gives
\begin{eqnarray}
    p &=& \frac{\mu^2x^2}{4R_{\rm in}^2(R_{\rm in}+2\mu)^2}+O(x^3),\nonumber\\
    \rho+p &=& \frac{\mu\,x}{2R_{\rm in}(R_{\rm in}+2\mu)}+O(x^2).
    \label{eq:pinner}
\end{eqnarray}
At the inner surface, $e^{2\lambda_{\rm in}} = R_{\rm in}/(R_{\rm in}+2\mu)$, $e^{2\nu_{\rm in}} = 1-2M/R$ by Eq.~(\ref{eq:firstint}), and $\nu'_{\rm in} = -\mu/[R_{\rm in}(R_{\rm in}+2\mu)]$. To leading order ${\cal P} = P_2^{(-)}x^2$, ${\cal W} = W_1^{(-)}x$, and ${\cal Q} = Q_1^{(-)}x$, with
\begin{eqnarray}
    P_2^{(-)} &=& \frac{\mu^2 e^{\lambda_{\rm in}+3\nu_{\rm in}}}{2R_{\rm in}^4(R_{\rm in}+2\mu)^2},\nonumber\\
    W_1^{(-)} &=& \frac{\mu e^{3\lambda_{\rm in}+\nu_{\rm in}}}{2R_{\rm in}^3(R_{\rm in}+2\mu)},\nonumber\\
    Q_1^{(-)} &=& -\frac{\mu^2(4R_{\rm in}+7\mu)e^{\lambda_{\rm in}+3\nu_{\rm in}}}{2R_{\rm in}^5(R_{\rm in}+2\mu)^3}.
    \label{eq:Q1inner}
\end{eqnarray}
Note that $Q_1^{(-)}<0$: because $\nu'_{\rm in}<0$, the term $4\nu'/r$ in ${\cal Q}$ dominates with a negative sign near the singularity, in contrast with $Q_1^{(+)}>0$ at the outer surface. The regular solution is $\zeta = \zeta_{\rm in}+\zeta_1^{(-)}x+O(x^2)$, $\eta = \eta_2^{(-)}x^2+O(x^3)$, with
\begin{eqnarray}
    \eta_2^{(-)} &=& -\frac{1}{2}\left(Q_1^{(-)}+\omega^2W_1^{(-)}\right)\zeta_{\rm in} ,\nonumber\\
    \zeta_1^{(-)} &=& -\frac{Q_1^{(-)}+\omega^2W_1^{(-)}}{2P_2^{(-)}}\zeta_{\rm in} .
    \label{eq:zeta1inner}
\end{eqnarray}
Using $Q_1^{(-)}/P_2^{(-)} = -(4R_{\rm in}+7\mu)/[R_{\rm in}(R_{\rm in}+2\mu)]$ and $W_1^{(-)}/P_2^{(-)} = R_{\rm in}^2/[\mu(1-2M/R)]$, the regularity condition takes the compact form
\begin{equation}
    \frac{\zeta'(R_{\rm in})}{\zeta(R_{\rm in})} = \frac{1}{2}\left[\frac{4R_{\rm in}+7\mu}{R_{\rm in}(R_{\rm in}+2\mu)}-\frac{\omega^2R_{\rm in}^2}{\mu(1-2M/R)}\right].
    \label{eq:inner_regular}
\end{equation}

To leading order near $R_{\rm in}$, the pulsation equation is $(P_2^{(-)}x^2\zeta')'+(Q_1^{(-)}+\omega^2W_1^{(-)})x\zeta \simeq0$, whose two independent local behaviors are $\zeta_{\rm reg}\sim{\rm const}$ and $\zeta_{\rm sing}\sim x^{-1}$. For the singular solution, $\eta={\cal P}\zeta'$ approaches a finite nonzero constant as $x\to0$. The free-boundary condition $\eta(R_{\rm in})=0$ therefore eliminates the singular branch, so the free-boundary and regularity conditions are equivalent. The same holds at the outer surface. At either endpoint, let $y$ denote the inward distance from the surface; then ${\cal W}\sim W_1 y$ with $W_1>0$. In the Sturm--Liouville norm, the regular local solution is square-integrable, $\int_0^\epsilon {\cal W}|\zeta_{\rm reg}|^2\,dy<\infty$, whereas the singular solution is not:
\begin{equation}
\int_0^\epsilon {\cal W}|\zeta_{\rm sing}|^2\,dy
\sim W_1\int_0^\epsilon\frac{dy}{y}=\infty\,.
\end{equation}
Thus each free-surface endpoint is of limit-point type. No additional independent self-adjoint boundary condition is required there in the weighted space of finite-norm radial displacements; the physical free-surface condition is realized by the admissible local branch.

%---------------------------------------------------------------------
%---------------------------------------------------------------------
\section{Horizon-penetrating evolution code}
\label{app:ef}

\emph{Field equations.} For the metric~(\ref{eq:efmetric}), the inverse metric in the $(v,r)$ block has $g^{vv}=0$, $g^{vr}=e^{-\beta}$, and $g^{rr}=F$, so $m$ is the Misner--Sharp mass. In spherical symmetry $\nabla_am=4\pi r^2\left(T_a{}^b\nabla_br-T^{(2)}\nabla_ar\right)$, where $T^{(2)}$ is the trace of $T_a{}^b$ over the $(v,r)$ block, which gives
\begin{equation}
\partial_rm=-4\pi r^2T^v{}_v\,,\qquad \partial_vm=4\pi r^2T_v{}^r\,.
\end{equation}
Since $g_{rr}=0$, $G_{rr}=-(2/r)\nabla_r\nabla_rr=(2/r)\Gamma^r_{rr}=2\partial_r\beta/r$, so $\partial_r\beta=4\pi rT_{rr}$.

\emph{Fluid variables.} For a perfect fluid $T_{ab}=(e+p)u_au_b+pg_{ab}$. The four-velocity has $u_r=e^\beta u^v\equiv z>0$, normalization gives $u_v=-e^\beta(1+Fz^2)/(2z)$, and $U\equiv u^r=(Fz^2-1)/(2z)$ is the Misner--Sharp velocity $dR/d\tau$. Equivalently, $z=1/(\tilde\Gamma-U)$ with $\tilde\Gamma^2=1+U^2-2m/r$; since $\tilde\Gamma>0$ for a collapsing shell, $z$ remains positive and finite inside trapped regions. A fluid element moves according to $dR/dv=u^r/u^v=e^\beta U/z$. With $D\equiv\rho_oz$, $Y\equiv(e+p)z^2$, and $X\equiv-T^v{}_v=\frac{1}{2}(e+p)(1+Fz^2)-p$,
\begin{equation}
\partial_r\beta=4\pi rY\,,\qquad \partial_rm=4\pi r^2X\,,
\end{equation}
and baryon conservation, $\partial_a(\sqrt{-g}\rho_ou^a)=0$ with $\sqrt{-g}=e^\beta r^2\sin\theta$, shows that the baryon mass between two fluid worldlines on a slice, $\Delta A=4\pi\int r^2D\,dr$, is constant.

\emph{Momentum equation.} Integrating $\partial_a(\sqrt{-g}\,T^a{}_r)=\frac{1}{2}\sqrt{-g}\,T^{ac}\partial_rg_{ac}$ over a zone whose edges move with the fluid, the advective fluxes cancel and only the pressure flux survives. Evaluating the source with $\partial_r\beta=4\pi rY$ and $\partial_rF=2m/r^2-8\pi rX$ gives Eq.~(\ref{eq:efmomentum}). For a static configuration, $z=F^{-1/2}$ and Eq.~(\ref{eq:efmomentum}) reduces to $p'=-(e+p)(m+4\pi r^3p)/(r^2F)$. Similarly, $dm/dv$ along a fluid worldline is $\partial_vm+(dR/dv)\partial_rm=-4\pi R^2p\,dR/dv$.

\emph{Thermodynamics and velocity recovery.} We evolve the adiabat $K$ of each zone, $p=K\rho_o^2$, which changes only through viscous heating, $dK/d\tau=q\,\dot\rho_o/\rho_o^3$. For $\Gamma=2$, $e+p=\rho_o+2p$, so $Y=Dz+2KD^2$, and $z=(Y-2KD^2)/D$ follows linearly from the evolved momentum, the geometry, and $K$.

\emph{Discretization.} Edges carry $R$ and the momentum $S=\int r^2Y\,dr$ of a dual cell extending between neighboring zone centers; zones carry $\Delta A$, $\Delta m$, and $K$. The mass function on the edges is $m_i=-\mu+\sum_{j<i}\Delta m_j$, and $\Delta m_j$ is updated by the work done on the zone edges, with the edge pressure the average of the adjacent zone pressures and zero at the surfaces. The dual-cell baryon masses are taken from the exact equilibrium profile: because $\rho_o\propto(\text{distance})$ near a surface, the half-zone adjacent to a surface contains about a quarter, not half, of the zone's baryon mass. With this choice the residual of the discrete equilibrium converges as $N^{-1}$ at the surfaces as well as in the interior. $\beta$ is integrated inward from $\beta=0$ at the outer surface using zone values of $Y$. The artificial viscosity is $q=(e+p)\left[\kappa_q^2(\Delta U)^2+\kappa_\ell c_s|\Delta U|\right]$ in compressing zones ($\Delta U<0$), with $\kappa_q=2$ and $\kappa_\ell=0.3$, and we integrate in time with the third-order strong-stability-preserving Runge--Kutta method. The time step is the smaller of an acoustic limit, based on the speed $e^\beta c_s/[z^2(1-c_s)]$ of ingoing sound waves relative to the fluid, and a source-term limit that restricts the relative change of $z$ and of each zone width per step to $5\%$; the latter controls the stiff repulsion $\propto\mu/r^2$ near the singularity.

\emph{Event horizon.} We store $\beta$ and $m$ on the grid every $0.1M$ in advanced time ($0.05M$ for $C=0.45$) and integrate outgoing null rays, $dr/dv=\frac{1}{2}e^\beta F$, with fourth-order Runge--Kutta, interpolating linearly in $r$ and $v$, and using $\beta=\beta(R_{\rm in})$, $F=1+2\mu/r$ in the cavity $r<R_{\rm in}$ and $\beta=0$, $F=1-2M/r$ outside the shell.

\bibliography{biblio}

\end{document}